# Theoretical and experimental study on Noise Equivalent Power of X-ray semiconductor ultra-fast response material based on the rad-optic effect

Xin Yan[a,b], Tao Wang[b], Gang Wang[b], Dong Yao[a,b], Yiheng Liu, Guilong Gao[b], Liwei Xin[a,b], Fei Yin[b], Jinshou Tian[b], Xinlong Chang[a*], Kai He[b*]

a *Rocket Force University of Engineering, Xi'an 710025, China*

b *Key Laboratory of Ultra-fast Photoelectric Diagnostics Technology, Xi'an Institute of Optics and Precision Mechanics (XIOPM), Chinese Academy of Sciences (CAS), Xi'an, Shaanxi 710119, China*

\* E-mail addresses: hekai@opt.ac.cn and xinlongch@vip.sina.com

**ABSTRACT**

Semiconductor material based on the rad-optic effect enables ultra-fast detection of X-rays and plays an important role in fusion diagnostics. Obtaining the accurate noise equivalent power (NEP) of the semiconductor ultrafast response material is the key to detecting X-rays. In this paper, the refractive index change mechanism of the semiconductor under X-ray irradiation was analyzed, and the quantitative relationship between the diffraction efficiency and the X-ray photon energy was established through the LT-AlGaAs diffraction imaging experiments. The impulse responses of LT-AlGaAs under 1 KeV-10 KeV X-ray radiation were calculated, revealing the variation of NEP density with radiated photon energy. In the case of bombarding the Al target to generate 1.5 KeV X-rays, the imaging experiments of LT-AlGaAs were performed. The diffraction image of LT-AlGaAs has a linear relationship with the radiation intensity, and the NEP density of LT-AlGaAs reaches $4.80\times10^5$ W/cm$^2$. This study has reference significance for the development of ultra-fast X-ray imaging systems based on the rad-optic effect.

effect, inertial confinement fusion(ICF)

## 1. INTRODUCTION

In the research of inertial confinement fusion (ICF), Multiple high-power laser beams are injected into the black cavity to generate X-rays, which heat and compress the target to form high-temperature and high-density plasma, and then achieve fusion. The duration of the critical process is generally less than 100 ps. X-rays of 1 KeV-10 KeV were generated by the black cavity with a high Z element under multiple high-power laser injections [1][2]. Additionally, in ICF studies, X-rays in the keV energy region generated by the interaction between the laser and the target are generally used as the backlight to conduct relevant diagnostic studies [3][4], so the diagnosis and the study of X-rays in the keV energy region are very important. By analyzing and measuring this X-ray radiation, a large amount of useful information related to the plasma state can be obtained [5], which is helpful for the in-depth study of important physical processes such as hot spot mixing, implosion-driven asymmetry, and hydrodynamic instability [6]. This can also provide a reference for the applied research on matching X-ray backlight in implosion physics, radiation opacity, and high energy density physics [3].

The semiconductors with the rad-optic effect are the important materials for achieving ultra-fast X-ray detection [7][8][9][10]. The Lawrence Livermore National Laboratory (LLNL) proposed ultra-fast detection technology based on semiconductor rad-optic effect, with time response up to 100 fs and advantages of resistance to electromagnetic radiation [11][12]. Richard E. Stewart et al. theoretically analyzed the minimum detectable energy of semiconductors such as GaAs and CdZnTe [13], and gave the minimum detectable power energy of 20 μm GaAs at $10^{-5}$ diffraction efficiency, which was about $2.6\times10^{-6}$ J/cm$^2$@16KeV. R. A. London et al. conducted a theoretical analysis and numerical calculation on the space-time characteristics of the energy deposition of gallium arsenide (GaAs) and evaluated the sensitivity of the system [14]. The Northwest Institute of Nuclear Technology conducted theoretical and experimental studies on an all-optical high-speed camera based on indium phosphating (InP). The time resolution and spatial resolution were about 1.5 ns and 140 μm respectively, and the excitation optical power density corresponding to the sensitivity of the system was about $1.3\times10^5$ W/cm$^2$ [15]. In recent years, the Xi'an Institute of Optics and Precision Mechanics, Chinese Academy of Sciences, has conducted theoretical and experimental studies on the photorefractive effect of LT-AlGaAs [16]. A low-temperature-grown (LT) GaAs/AlGaAs multi-quantum well (MQW) device was used as the response material of an imaging system to obtain the six-amplitude imaging results with a time resolution of 3 ps [17]. However, X-ray imaging experiment and spectral response analyses of LT-AlGaAs were not carried out.

In this research, the refractive index change of semiconductor materials under radiation was calculated by analyzing the band gap contraction effect, band filling effect, and free carrier absorption effect. The change of the diffraction light intensity caused by the refractive index change was calculated and the variation of NEP density with radiated photon energy was revealed. The NEP density of LT-AlGaAs was experimentally studied by using the X-rays generated by a nanosecond laser as the signal. The experimental results verify the feasibility of high SNR X-ray imaging using LT-AlGaAs. This study provides theoretical and experimental support for the response analysis of the ultra-fast X-ray imaging system based on the rad-optic effect.

## 2. PHYSICAL MODEL

The spectral response of semiconductor materials is analyzed based on the method of transient grating diffraction imaging. As shown in Fig. 1, a metal grating mask is designed on the surface of semiconductor material. When the X-ray radiation pulse is incident on the semiconductor material, the transient refractive index change corresponding to the mask is formed inside the material, and then the phase grating is formed. The probe beam incident on the other side is reflected by the surface, forming a phase change consistent with the transient refractive index change. After diffraction by the phase grating, a diffraction image with the radiation pulse information can be obtained. Therefore, the ultra-fast response characteristics of semiconductor materials with different incident photon energies can be analyzed through the intensity of the diffraction images.

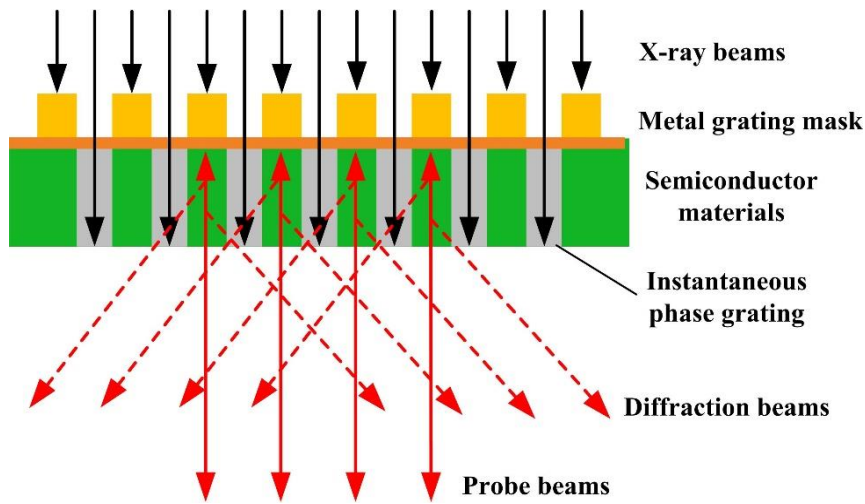

**Fig. 1 Grating diffraction imaging principle of photorefractive materials**

The number of electron-hole pairs produced by X-ray has no obvious relationship with the type of radiation, but rather is proportional to the total energy deposited $W$ by radiation:

$$\epsilon = \frac{W}{N}, \tag{1}$$

where $\epsilon$ is the average amount of energy required to produce one electron-hole pair, which varies with the type of semiconductor and is about three times the band gap width.

The intensity of an X-ray in a semiconductor material decays exponentially, which can be written as:

$$I_x(x,t) = \exp(-\alpha x) \cdot I_x(t), \tag{2}$$

where $x$ is the depth of the semiconductor material, $\alpha$ is the photoelectric absorption coefficient of the material, and $I_x(t)$ is the X-ray intensity at time $t$.

When the carrier recombination is ignored, the excess carrier concentration is a function of the depth:

$$N(x) = \frac{\alpha}{\hbar\omega}\exp(-\alpha x)\int_{-\infty}^{\infty} I_x(t)dt. \tag{3}$$

The generated excess carrier concentration $N(x)$ causes the change of the material refractive index. By calculating the change of the material absorption rate $\Delta\alpha(N,E)$ caused by excess carriers, the change of the refractive index can be obtained using the Kramers-Kronig change relationship. The relationship between the change of the refractive index $\Delta n(N,E)$ and the change of the absorption rate $\Delta\alpha(N,E)$ is as follows:

$$\Delta n(N,E) = \frac{c\hbar}{\pi} P \int_0^{\infty} \frac{\Delta\alpha(N,E')}{E'^2 - E^2} dE', \tag{4}$$

where $N$ is the excess carrier concentration, $E$ is the photon energy of different energy spectra, and $P$ is the Cauchy principal value.

When the probe beam is incident, the phase shift caused by the refractive index change can be described as follows:

$$\Delta\Phi(t) = \frac{4\pi}{\lambda} \int_0^d \Delta n(x) dx, \tag{5}$$

where $d$ is the thickness of the refractive index change of the material. For the case in which the absorption depth is smaller than the material thickness T, $d = \frac{1}{\alpha}$. For the case in which the absorption depth is larger than the material thickness, $d = T$.

The diffraction efficiency of the ±1st order grating used for imaging can be expressed as:

$$\varepsilon(1) = \varepsilon(-1) = \frac{4}{\pi^2} \sin^2(\Delta\Phi). \tag{6}$$

Assuming that the light intensity of the probe is $I_{Probe}$, the diffraction light intensity passing the +1 or −1 level is proportional to the diffraction efficiency and is a function of refractive index change $\Delta n$:

$$I = I_{Probe} \cdot \frac{4}{\pi^2} \sin^2\left(\frac{4\pi}{\lambda} \int_0^d \Delta n(x) dx\right). \tag{7}$$

The refractive index changes are mainly caused by the band gap contraction effect, band filling effect, and free carrier absorption effect. The refractive index changes caused by three effects are analyzed to calculate the light intensity of the diffraction image, and then the spectral response characteristics of the photorefractive semiconductor materials are obtained.

In the excitation process, the semiconductor valence band electrons absorb photon energy and transition to the conduction band to produce non-equilibrium carriers, resulting in the partial energy levels of the valence band being occupied by holes and the partial energy levels of the conduction band being occupied by electrons to reduce the carrier distribution function in each energy band. The phenomenon of the change of the distribution function caused by carriers is the band filling effect (BF).

The absorption coefficient of semiconductor materials is given by the square root law:

$$\begin{aligned}\alpha_0(E) &= \frac{C}{E}\sqrt{E-E_g} & E \geq E_g \\ \alpha_0(E) &= 0 & E < E_g\end{aligned} \quad (8)$$

where $E = h\nu$ is the excited photon energy, $E_g$ is the band gap energy of the material, and $C$ is a constant related to the material. Considering the effect of light and heavy holes on absorption, Equation (6) can be written as

$$\begin{aligned}\alpha_0(E) &= \frac{C_{hh}}{E}\sqrt{E-E_g} + \frac{C_{lh}}{E}\sqrt{E-E_g} & E \geq E_g \\ \alpha_0(E) &= 0 & E < E_g\end{aligned} \quad (9)$$

where $C_{hh}$ and $C_{lh}$ correspond to heavy holes and light holes, respectively.

In the band filling effect, the state distribution of the conduction band occupied by an electron or the valence band of a vacant electron follows the Fermi-Dirac distribution. Assuming that the energy level of the electron in the valence band is $E_a$ and the energy in the conduction band is $E_b$, the absorption coefficient of the semiconductor material after carrier injection can be written as follows:

$$\alpha(N_e, N_h, E) = \alpha_0(E)[f_v(E_a) - f_c(E_b)], \quad (10)$$

where $N_e$ and $N_h$ represent the concentrations of free electrons and holes, respectively, and $\alpha_0$ represents the absorption coefficient of the intrinsic materials. $f_v(E_a)$ represents the probability that electrons occupy the valence band energy level $E_a$, and $f_c(E_b)$ represents the probability that electrons occupy the conduction band energy level $E_b$. Owing to the valence band degeneracy, each energy level has two values. $E_{ah}$ and $E_{al}$ correspond to the energy levels of the heavy and light holes in the valence band, and $E_{bh}$ and $E_{bl}$ correspond to the energy levels of the heavy and light holes in the conduction band. Then the change of the absorption coefficient of the semiconductor material caused by the band filling effect is

$$\Delta\alpha(N_e, N_h, E) = \alpha(N_e, N_h, E) - \alpha_0(E)$$
$$= \frac{C_{hh}}{E}\sqrt{E - E_g}[f_v(E_{ah}) - f_c(E_{bh}) - 1]$$
$$+ \frac{C_{lh}}{E}\sqrt{E - E_g}[f_v(E_{al}) - f_c(E_{bl}) - 1] \quad (11)$$

Using the Kramers-Kronig relationship, the change of the refractive index $\Delta n_F$ caused by the band filling effect can be obtained:

$$\Delta n_{BF} = \frac{c\hbar}{\pi} P \int_0^\infty \frac{\Delta\alpha_F\left(\Delta N_e(z,t), \Delta N_h(z,t), E'\right)}{E'^2 - E^2} dE' . \quad (12)$$

When the concentration of free carriers in the materials reaches a certain value, the band gaps decrease as the concentration of carriers continues to increase, a phenomenon known as band gap shrinkage (BGS). Equation (13) gives the calculation formula of the band gap change:

$$\Delta E_g = \frac{\kappa}{\varepsilon_s}\left(\frac{N}{N_{CR}} - 1\right)^{\frac{1}{3}}, \quad (13)$$

where $\kappa$ is a constant, $\varepsilon_s$ is the static dielectric constant, and $N_{CR}$ is the critical free carrier concentration.

When the energy band is occupied by unbalanced carriers, the probability of exciting radiation is greater than the absorption probability, which affects the optical properties of the material. The absorption rate of the semiconductor material caused by this can be expressed as:

$$\alpha = C_\alpha \frac{\sqrt{E - E_g}}{E}[f_1(E) - f_2(E)]. \quad (14)$$

The final change in the absorption rate caused by the band gap contraction is as follows:

$$\Delta\alpha_{BGS} = \frac{C_\alpha}{E}\left(\sqrt{E - E_g + \kappa\left(\frac{N}{N_{CR}} - 1\right)^{\frac{1}{3}}} - \sqrt{E - E_g}\right). \tag{15}$$

Using the Kramers-Kronig relationship, the change of the refractive index $\Delta n_S$ caused by the band gap contraction can be obtained as follows:

$$\Delta n_{BGS} = \frac{c\hbar}{\pi} P \int_0^\infty \frac{\Delta\alpha_S\left(\Delta N_e(z,t), \Delta N_h(z,t), E'\right)}{E'^2 - E^2} dE'. \tag{16}$$

Free carrier absorption (FCA) is the change of refractive index $\Delta n_A$ due to the interaction between the incident probe beam and the free carriers, which can be described by the Drube plasma absorption relationship:

$$\Delta n_{FCA} = -\left(\frac{e^2\lambda^2}{8\pi^2 c^2 \varepsilon_0 n}\right)\left(\frac{\Delta N_e(z,t)}{m_e} + \frac{\Delta N_h(z,t)}{m_h}\right), \tag{17}$$

where $\lambda$ is the wavelength of the probe beam, $m_e$ is the effective mass of the electron, $m_h$ is the effective mass of the hole, c is the speed of light in a vacuum, and $\varepsilon_0$ is the dielectric constant of a vacuum.

Considering the comprehensive influence of the three effects, assuming that the three effects are independent of each other, the refractive index change is the sum of the refractive index change caused by the three effects.

$$\begin{aligned}\Delta n &= \Delta n_F + \Delta n_S + \Delta n_A \\ &= \frac{c\hbar}{\pi} P \int_0^\infty \frac{\Delta\alpha_{BF}\left(\Delta N_e(z,t), \Delta N_h(z,t), E'\right)}{E'^2 - E^2} dE' \\ &+ \frac{c\hbar}{\pi} P \int_0^\infty \frac{\Delta\alpha_{BGS}\left(\Delta N_e(z,t), \Delta N_h(z,t), E'\right)}{E'^2 - E^2} dE'. \\ &- \left(\frac{e^2\lambda^2}{8\pi^2 c^2 \varepsilon_0 n}\right)\left(\frac{\Delta N_e(z,t)}{m_e} + \frac{\Delta N_h(z,t)}{m_h}\right)\end{aligned} \tag{18}$$

Through Equations (7) and (18), the light intensity of the diffraction images obtained with the +1 or −1 levels can be obtained.

In this research, 10 μm LT-AlGaAs was used as an ultra-fast response material to analyze the spectral response. To ensure the light reflection of the probe, 100 nm Al was used between the metal grating and the semiconductor material to form a high reflection film of the probe beam, and the photon energy transmittance between 1 keV and 10 keV was greater than 90%. Therefore, the influence of the high reflection film on the spectral response was not considered.

LT-AlGaAs absorption coefficient $\alpha$ is a function of photon energy $E$. The absorption depth $\frac{1}{\alpha(E)}$ at different photon energies is shown in Fig. 2. It can be seen that the incident photons in the range of 1 KeV–5KeV can be completely absorbed by a 10 μm material, while the incident photons with higher energy cannot be completely absorbed.

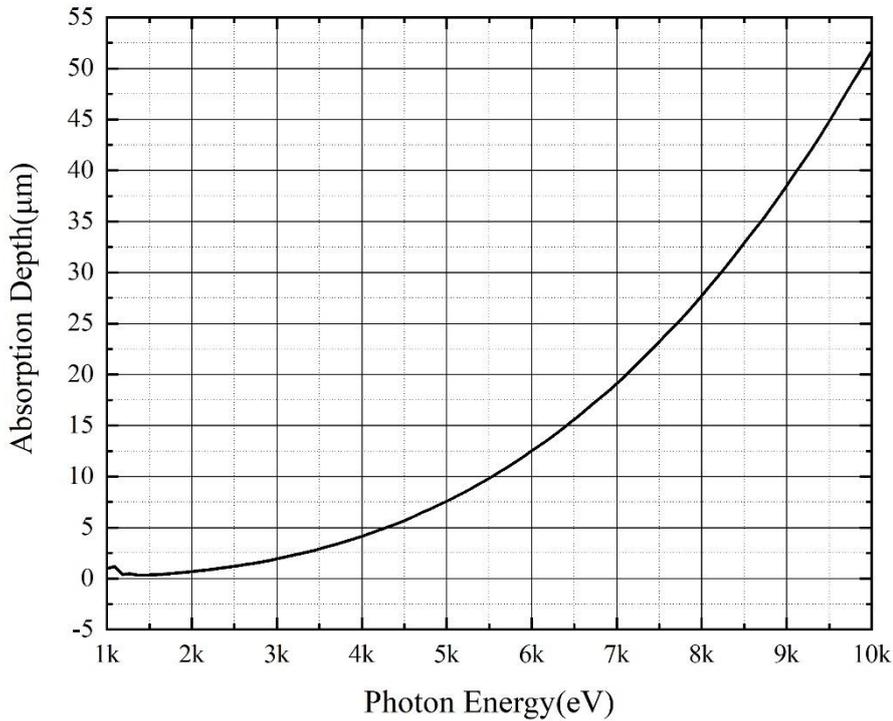

**Fig. 2 Absorption depth of LT-AlGaAs at different photon energies**

Generally, a laser that matches the LT-AlGaAs gap width is used as the probe beam. When a 780 nm probe beam is used, the diffraction efficiencies of different radiation intensities within the range of 1 KeV to 10 KeV are as shown in Fig. 3.

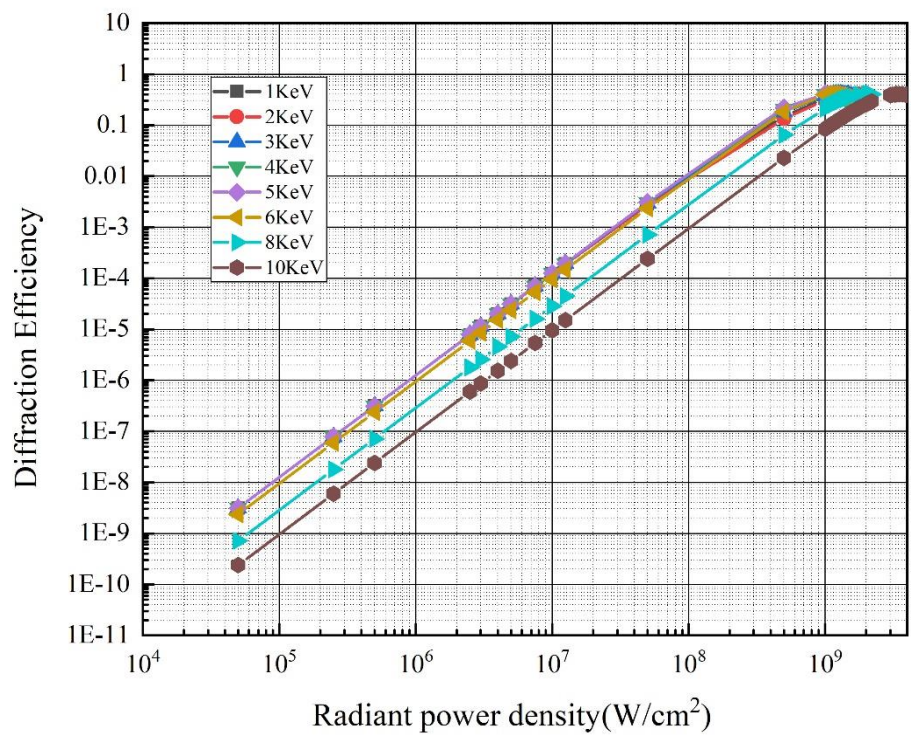

（a）Relationship between diffraction efficiency and photon energy at different radiation power densities

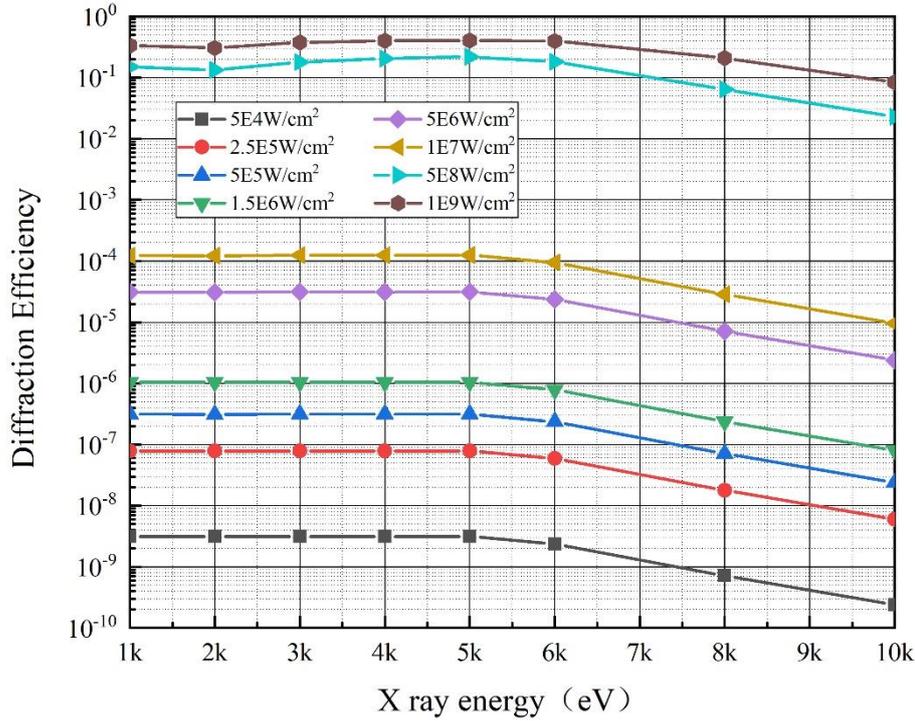

(b) Relationship between diffraction efficiency and radiation intensities density at different photon energies

**Fig. 3 Diffraction efficiency of LT-AlGaAs with different radiation intensities in the range of 1 KeV to 10 KeV**

It can be found that when the thickness of semiconductor material exceeds the depth of X-ray absorption, the X-ray photon deposits energy in the semiconductor, and the diffraction efficiency is related only to the material parameters, not the X-ray photon energy. When the thickness falls below the depth of X-ray absorption, the X-ray photon energy deposited into the semiconductor material diminishes and the diffraction efficiency of the chip declines with the increase of X-ray photon energy. As shown in Fig. 2, nearly all of the photons below 6 keV are absorbed by the 10 μm AlGaAs, as a result of which the curves in the range of 1 KeV to 5 KeV overlap. In the energy segment of 6 KeV to 10 KeV, the depth of absorption exceeds 10 μm with the increase of incident photon energy, which reduces the level of diffraction efficiency.

To improve the efficiency of detection for high-energy X-rays, it is suggested that a thicker response layer of ultra-fast semiconductor material can be developed in the future to enhance the deposition efficiency of high-energy X-ray.

## 3. EXPERIMENT and ANALYSIS

### 3.1 Experimental layout

A nanosecond pulse laser was used to bombard an aluminum (Al) target for producing X-ray pulses. The laser energy ranged from 5 J to 155 J, and the width of the laser approached 2 ns. The response layer of LT-AlGaAs was 10μm, and the 780 nm laser was treated as a probe laser. As shown in Fig. 4, the distance between the semiconductor and the metal target was about 7 cm. Andor intensified charge-coupled device (ICCD) was employed to achieve synchronization with the target laser, with a gain of 3000, and a gate width of 150 ns. A 2μm Al film was placed in front of the semiconductor to shield scattering light in the target chamber. When X-ray is incident on the semiconductor mask surface, the changes of periodic refractive index were formed in LT-AlGaAs material, and the probe light was modulated by the phase grating inside the semiconductor. A spatial filter was fixated on the spectrum surface of the 4f imaging system to allow only +1 order signal light to pass through and block the stray light of other frequencies, thus improving the signal-to-noise ratio of the imaging system.

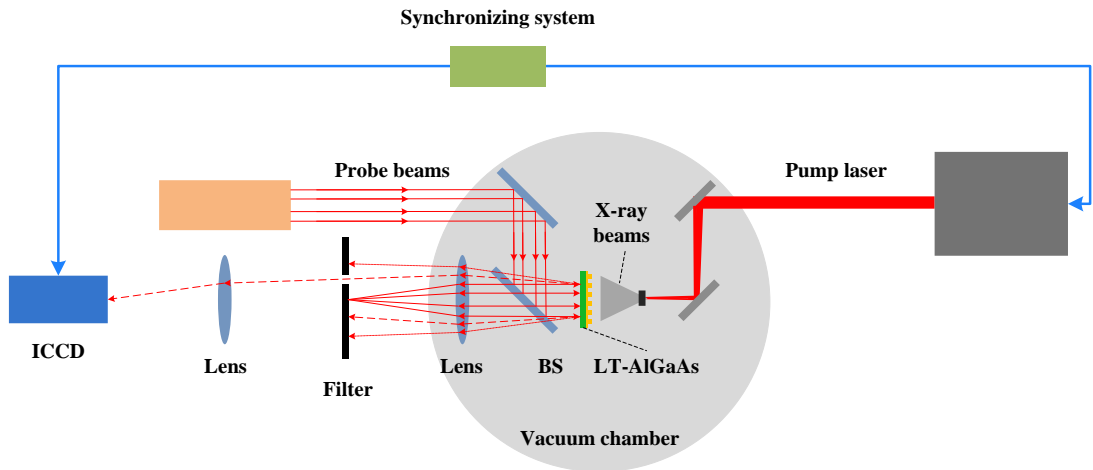

**Fig. 4 Layout of the experimental system**

3.2 Results

In the experiment, the response characteristics of the system to different radiation intensities were obtained by adjusting the pulse energy of the target laser. The X-ray imaging results are shown in Fig.5. The gray values of pixels in the image region obtained with ICCD were averaged to obtain the average gray values of the single pixels. The laser pulse energies were 155 J, 45 J, 23 J, and 5 J. The X-ray power density measured by XRD was converted to the chip surface as follows: $3.14\times10^7$ W/cm$^2$, $1.51\times10^7$ W/cm$^2$, $6.12\times10^6$ W/cm$^2$, and $4.91\times10^5$ W/cm$^2$.

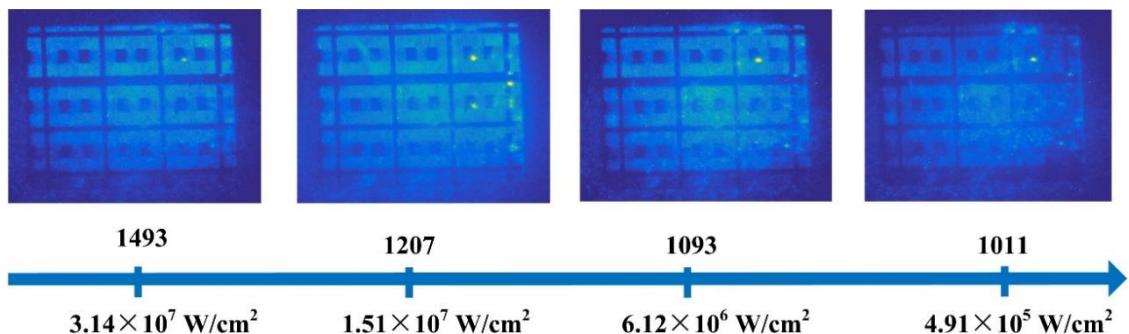

**Fig. 5 Average gray value of a single pixel for imaging for different radiation fluxes**

For the different X-ray radiation intensities adopted in the experiment, the system could image well and display the details of the target. With the enhancement of the radiation intensity, the overall gray value of the imaging also increased. In this system, the minimum detectable X-ray power density was better than $4.91\times10^5$ W/cm$^2$. Some of the bright spots were mainly caused by strong scattering light that was in turn caused by defects in the semiconductor manufacturing process. The image uniformity could be improved by optimizing the manufacturing process.

The characteristic X-ray produced by high-energy laser bombarding the Al target is 1.5 KeV. As can be seen from Fig.2, the 10 μm AlGaAs used in the experiment can completely absorb the X-ray of 1.5 KeV.

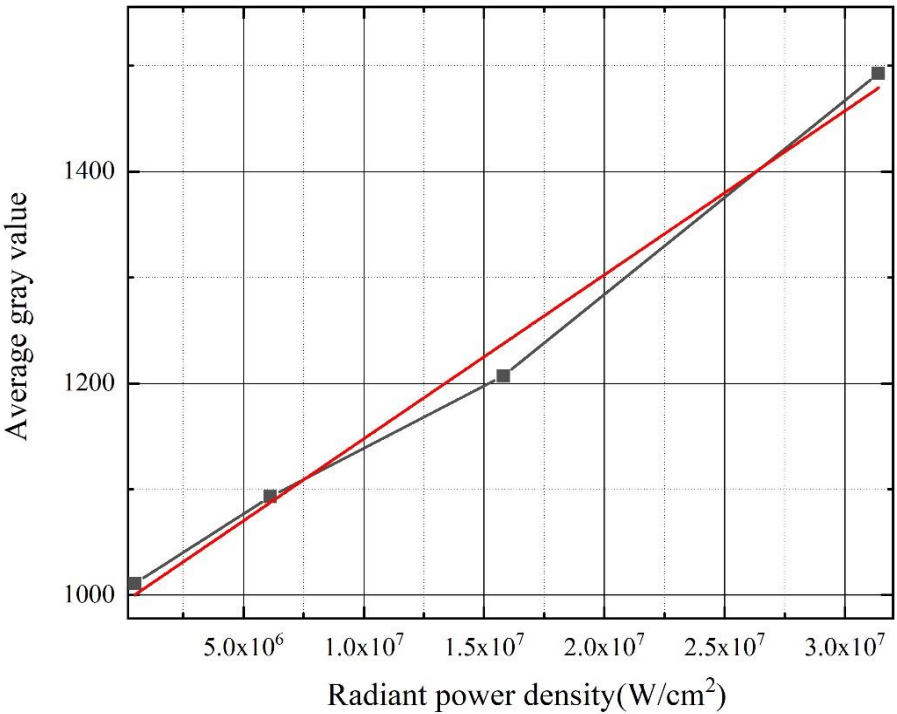

**Fig. 6 Experimental results of diffraction signal intensity under different X-ray power density**

Fig.6 shows the experimental results of diffraction signal intensity under different X-ray power densities. Combined with Fig.3 (a), when the X-ray radiation power density is $4.91 \times 10^5$ W/cm$^2$, the diffraction efficiency of the semiconductor chip is $3.0 \times 10^{-7}$. The red line in the figure linearly fits the experimental data, and the R-square value is 0.995, indicating that the average gray value of a single pixel met the linear relationship for the condition of the experimental X-ray intensity. Through fitting calculation, the NEP density of the semiconductor used in this experiment is 4.80×10⁵ W/cm².

The noise of the system is mainly caused by the scattering of probe light from the rough surface of the semiconductor. The scattered light in the same direction as the diffraction signal light was collected by ICCD through the 4f imaging system. Related to semiconductor processing, this part of noise can be reduced by optimizing the process of substrate removal.

The semiconductor used in the experiment has a surface roughness of about 10 nm. The random height variation of the semiconductor surface causes incident light to be scattered. In general, the scattering characteristics of material surface can be indicated by the Bidirectional Scattering Distribution Function (BSDF), which is defined as the ratio of scattering brightness to incident illumination. Reference can be made to [18] for the following calculations.

$$BSDF(|\sin\theta_s - \sin\theta_i|) = b_0 \left[1 + \left(\frac{|\sin\theta_s - \sin\theta_i|}{l}\right)^2\right]^{s/2}$$

where $b_0 = \frac{4\pi^2 \Delta n^2 QA}{\lambda^4}$, $l = \frac{\lambda}{B}$, $s = -C$, $\theta_i$ represent the incident angle, $\theta_s$ indicates the scattering angle, $\lambda$ denotes the wavelength of incident light, $\Delta n$ refers to the difference of refractive index between the scattering surface, Q represents the polarization factor, A denotes the magnitude of spectral power density at low frequency, $1/B$ indicates the spatial frequency of rolling, and C refers to the slope of spectral power density when the spatial frequency exceeds $1/B$. The probe light is incident vertically. The roughness of chip surface is 10 nm, which is far smaller than the probe light wavelength, so that $\Delta n$=2, Q=1. The parameters of k-correlation model are as follows: $A = 4.64 \times 10^{-3} \mu m^4$, $1/B = 1 \times 10^{-3} \mu m^{-1}$, $C = 1.55$. The aperture diameter of holes is 3 mm.

The scattered light received by ICCD accounts for 1.1×10⁻⁷ of the intensity of incident light. As can be seen in Fig.3(a), the corresponding noise equivalent power density is around $3.0\times10^5$ W/cm$^2$. The experimental results are found consistent with the calculation results, which verify the theoretical calculation of characteristics of LT-AlGaAs response in the range of 1 KeV-10 KeV. According to the definition of BSDF, the NEP density of the system can be lowered by optimizing the process of semiconductor preparation and reducing the aperture, to reduce the scattering rate.

Compared to the experimental data, the NEP density calculated by scattering efficiency is lower, for which the primary reason is the scattered light caused by other factors, such as the scattered light introduced by slight scratches, dents, and particulate pollutants on the semiconductor surface during the experiment.

## 5、CONCLUSION

The all-optical solid ultra-fast imaging technology based on the rad-optic effect is promising for extensive application in ultra-fast X-ray detection. In this paper, the NEP of LT-AlGaAs ultra-fast response material was studied, and the response characteristics of the imaging system with high SNR in the X-ray spectrum were verified. The response characteristics of materials to different photon energy were calculated, to reveal the changes in NEP density with radiated photon energy. With ultra-fast response LT-AlGaAs material, X-ray imaging experiments were carried out, and the experimental results were consistent with the calculation results. The NEP density of LT-AlGaAs reached $4.80\times10^5$ W/cm$^2$. The results of this study provide a reference for the application of ultra-fast all-optical solid imaging technology with rad-optic effect in X-ray detection.

**Acknowledgments**

This work was supported by the National Natural Science Foundation of China

(Grant Nos. 12075312 and 62005311).(Grant Nos. 12075312 and 62005311).